# Delving into the Security Issues of Mobile Cloud Computing

Navdeep Kaur
Dept. of Computer Science and Engineering
Guru Nanak Dev University
Amritsar, India
navdeep_binny@yahoo.com

Prabhsimran Singh
Dept. of Computer Science and Engineering
Guru Nanak Dev University
Amritsar, India
prabh_singh32@yahoo.com

*Abstract*— Looking at the last decade, progress in technology has made a huge impact on our lifestyles. Enhanced use of mobile phones has provided a technological breakthrough, with the latest smartphones capturing the market. The word smartphone is enough for everyone to understand the tremendous potential it brought to the market in terms of economics as well as usability. Not only this, this ever growing mobile mania has a lot more to offer. The familiarity of applications like dropbox etc is a clear indication of the popularity of mobile and cloud computing. But where we get all the benefits from this computing platform, there are some of the challenges too. However, with the enhanced facilities and luxuries, some challenges are always accompanied.

## I. INTRODUCTION

Mobile Cloud Computing is emerging as one of the main key areas of cloud computing. Cloud computing is an added advantage as it adds to the functionality provided by mobiles in terms of storage, enhanced processing capability and a lot more. Mobile cloud computing can be viewed as cloud computing with portability. Using services of the cloud is economical to the user and generates revenue for the cloud provider. Therefore this concept on the whole is promoted in all the domains related to computers[1]. This paper explains threats in mobile cloud introduced at various layers of the cloud computing architecture-SaaS, PaaS and IaaS. A brief description of these threats has been given along with generalized measures that should be taken. Some security breaches at the network level and web application level have been discussed. Issues specific to the mobile domain are also presented.

## II. NEED OF SECURITY

The basic architecture of cloud computing is based upon three service delivery models viz IaaS, PaaS and SaaS, with IaaS being the foundation layer followed by PaaS and then further by SaaS. Trade offs in terms of extensibility and security always pertains to these layers. More and more companies are inclined towards adopting cloud computing at a larger scale in their organization but security issues related to it remain as a major hurdle. To guarantee the security of data in these three layers is difficult as each layer has its own security issues.

Security is required because of following reasons[1]:
    a. Transmission of personal data to the cloud server.
    b. Transmission of data from cloud server to clients computer.
    c. Storage of clients personal data in remote cloud servers which are not owned by the client.

Figure 1 shows infrastructure of mobile cloud with the basic operations that it needs to provide to all its customers. Security issues will be discussed keeping in view the services that the mobile cloud is expected to provide.

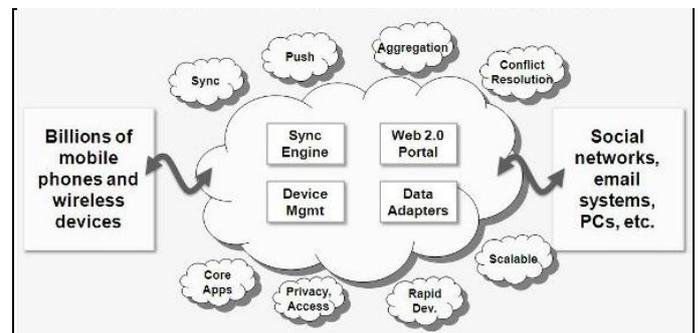

Figure 1, "Infrastructure of Mobile Cloud"

## III. LAYERS OF MCC

The 3 layers of Mobile cloud Computing are:

### A. SaaS

In SaaS, the client has to depend on the provider for proper security measures. The provider must do the work to keep multiple users' from seeing each other's data. So it becomes difficult to the user to ensure that right security measures are in place and also difficult to get assurance that the application will be available when needed. With SaaS, the cloud customer will by definition be substituting new software applications for old ones. Therefore, the focus is not upon portability of applications, but on preserving or enhancing the secure

functionality provided by the legacy application and achieving a successful data migration

i.   **Data Confidentiality:** Senstive data of the organization is accessible to the software services that are being provided by different vendors. Security and privacy policies of different cloud service providers may not be in tune with the confidentiality of the user. Maintaining the security of all governmental and non governmental organizations is critical to the success of cloud provider. Some of the threats at this level include: Cross-site scripting, Hidden field manipulation, cookie manipulation etc[3]. Spyware and malware poses another threat that can affect data centers if proper measures are not taken by the cloud providers.

ii.  **Data locality:** When sensitive information is involved, even intermediate data and its storage is of utmost importance. In case of data failure, if there are no measures for backup and recovery, it can lead to serious data loss. Location of data centers should be away from areas prone to natural disasters and data should be distributed so that entire data is not available to unauthorized person even in case of uncaught intrusion.

iii. **Multi-tenancy:** Data segregation is always there at the physical level provided by os level memory protection. However data of various users using a particular application can act as intruder for other users. Ensuring a clear boundary for each application is must in SaaS.

iv.  **Data access:** Now security policies of two organizations may vary. One may be more stringent or lenient than the other. In case of lenient security policies , coz data is also stored at the cloud server of some organization, data access may be not very secure.[3]

v.   **Authentication and Authorisation:** There may be a third party authentication server in between cloud server and the user or the authentication may be provided by the cloud provider itself. This process is transparent to the user.[2]Any request to access cloud must pass through this authentication server. This process of verification undoubtedly adds to the performance overhead but the process ensures that unauthenticated users are warded off at the earliest stage possible.

vi.  **Web application security:** Apart from the security risks at the network level, web applications introduce additional security threats. The Open Web Application Security Project has identified Top security risks faced by web applications. Those threats are:
   1. Injection flaws like SQL, OS and LDAP injection
   2. Cross-site scripting
   3. Broken authentication and session management
   4. Insecure direct object references
   5. Cross-site request forgery
   6. Insecure cryptographic storage
   7. Failure to restrict URL access
   8. Insufficient transport layer protection [2]

*B. PaaS*

It stands for Platform as a service. With PaaS a developer doesn't need to but all the tools. For example, GoogleApps, which is a collection of Google tools including, gmail, google drive, hangouts etc. two basic issues are of concern under this category.
   1. The security of the platform itself and
   2. Security of the customer applications that use this platform for storage or for running their applications.

Most of the issues related with the second point have been discussed in the above service model (SaaS)[4].

i.   **Third Party Relationship:** To support PaaS, a lot of interface level and component level support is needed to enable interoperability of various user applications and the underlying platform. MASHUPS are created using Ajax, by integrating elements from two or more sources. Data and network security issues related to mashups also arise here. Example, Google maps and Microsoft's Virtual earth have been integrated into a mash-up known as Flash Earth, which is zoomable. Any loophole in the components used can lead to serious security flaws.

ii.  **Different Versions of platform services(SDLC):** Now applications from every domain experience changes with the changes being made at organization, technological, requirement etc levels. To keep up with these changes, PaaS has to be upgraded and made flexible enough to allow different versions of these applications run smoothly. Such changes may compromise with the security of the cloud[4].

*C. IaaS*

It stands for Infrastructure as a service. IaaS provides a pool of resources such as servers, storage, networks, and other computing resources in the form of virtualized systems, which are accessed through the Internet. Example, Amazon's EC2.This cloud provider provides a virtual environment supporting Linux, Solaris or Windows. Users can run any software with full control and management on the resources allocated to them. However, the underlying compute, network, and storage infrastructure is controlled by cloud providers. IaaS providers must undertake a substantial effort to secure their systems in order to minimize these threats that result from creation, communication, monitoring, modification, and mobility. Here are some of the security issues associated to IaaS.[4]

i. **Virtualization:** Virtualization is yet another major component of cloud computing. To ensure isolation of different applications running on the virtual devices is a matter of concern. Virtual environments should be equally secure as physical environments since virtualization leads to addition of another layer in the cloud and attackers have yet another area to target. Various threats pertain due to the virtual environment such as VM escape, VM hopping and Malicious VM attacks etc.

ii. **Resource Sharing:** Virtual machines are designed in such a way that they share resources(CPU, I/O, memory) with each other. A malicious VM can obtain information about other VMs. Virtual Machine Monitor(VMM) is responsible for managing these virtual machines so that they work in complete isolation despite using shared resources. Flaws in designing VMM may lead to malicious machines accessing information easily and thus a weak security level.

## IV. SOME OTHER SECURITY ISSUES

The security issues are explained as following:

### A. Network Security

Since mobile devices are connected through a network, it forms one of the greatest vulnerability to security breach. Man in the middle attacks, DOS attacks, Ip sniffing and many more make network one of the most crucial areas while implementing security measures. Various encryption techniques have been in use since a long to protect data from intruders such as Secure Socket Layer(SSL) and Transport Layer Security(TLS)[3]. Network is the most vulnerable component when it comes to discussing security issues in cloud environment. Some of the network level threats are:

i. **Wifi Sniffing:** A number of hardware and software devices are available to act as wifi sniffer. Through wifi sniffing, anyone can monitor either the location of device or the activity being done. All wifi sniffing apps were removed by apple from its app store in march 2010.

ii. **Session Hijacking:** In computer science, session hijacking, sometimes also known as cookie hijacking is the exploitation of a valid computer session, sometimes also called a session key to gain unauthorized access to information or services in a computer system. The intruder hijacks a real time session and therefore it becomes difficult to detect. Now since mobile banking is widely used and promoted, this type of threat is one of the prominent ones and forms a major hurdle in increasing the user base of cloud computing[5].

### B. Web Based Threats

i. **Phishing scams:** Now since smartphones are there and all of the mails are available right there in ur cell phones, this increases threats to your mobile device. It is generally observed that a lot of people receive mails that falsely claim to be an established enterprise in an attempt to gain access to user's personal data. In 2003, such mails were received from bogus website claiming to be from ebay to update credit card information.

ii. **Drive by downloads:** Drive-by download means two things, each concerning the unintended download of computer software from the Internet. e.g. downloads which install an unknown or counterfeit executable program, ActiveX component, or Java applet. Hackers use this technique very cleverly by setting up websites with viruses. So whenever a user accesses those websites, unwanted programs get installed in the system. These sites can easily spread through the social media and infect other systems as well. This is why major companies keep on upgrading their OS to remove such security threats.

iii. **Browser exploits:** In this case, a malicious code can take advantage of a flaw or vulnerability in the OS to breach browser security and alter settings of the browser without user's consent. Any mobile device accessing a webpage through this browser is vulnerable to security threats.[5]

iv. **Jail broken devices:** Jail breaking is a famous method of accessing root directory of Apple's iOS platform. iPhones and iPads lose lots of their security through jail breaking[5]. Example, JailbreakMe is a series of browser based exploits used to jailbreak Apple's iOS mobile operating system. It uses an exploit in the browser's PDF parser to execute unauthorised code and gain access to the underlying operating system.

v. **Using Open Wifi:** Mobiles are more susceptible to the security threats offered by open wifi. Suppose that the wifi you are using is not encrypted. In such a case, hackers can easily get into that network and any of the above network level attacks can be easily done.

## V. MOBILE DOMAIN SPECIFIC ISSUES

Mobiles are prone to security threats at a greater level than computers while using cloud computing:

1. **Insecure Connectivity:** Suppose a person goes to a shopping mall, restaurant etc. Now since, mobile phones and forever connectivity has become a part and parcel of life, ensuring secure connectivity is mandatory. Every person may not always be carrying a laptop with him but a mobile phone is must, which supports the fact that mobile devices are much larger in number. Also, wifi though secured with passwords, are more susceptible to sniffing and other attacks, which is not as easy with cellular networks. Thus they are more prone to security threats.

2. **Quick access of critical data:** Emails and other applications are directly accessible through smart phone. A person using a smartphone doesn't necessarily have all the knowledge of these security threats. And equally sensitive and private data is present in mobiles as well as computers. Ensuring safe connectivity is even more critical.

3. **Web browsing for handheld devices:** Almost all modern mobile OS such as Android, IOS have the support for web browsing. This makes malware, spyware and other such threats to easily infect the mobile devices as users unknowingly click on these links while accessing websites on their phones.

4. **Excessive use of cloud resources without security:** Because mobiles have limited scalability, using different applications remotely via a cloud is a daily affair. The data within these applications may contain sensitive information and processing using cloud can be less secure if all security measures are not taken. Moreover, use of antivirus is still not widespread in mobile phones.

5. **Enhanced socializing:** Increasing social networking leads to increased exposure of your privacy to the outside world. And with all these sites available for the mobile phones, a person is always online. Malicious links available on these sites make the entry of Trojans and viruses easy for your mobile device and hack your bank details and passwords and send these to hackers.

6. **Location Services:** Tracking location has become quite easy using gps available in all the smartphones. This has further made crimes easy. Lack of privacy as well as security is the result of being tracked by location services all the time.

7. **Mobile Wallet:** With a lot of e-commerce gaining popularity and use of mobile banking on the rise, hackers are becoming keen to attack mobile devices to access personal data and hoard a lot of money. OTPs are also accessible directly through the apps once the device receives a message. This makes crime easier once access is gained to your device.

## VI. COUNTER-MEASURES

Some of the measures that can be adopted against these threats are as under[6]:

1. **Enhanced Security Policy:** Stringent security policies need to be implemented to keep a check on the malicious activity. Along with this, protocols need to be more modified keeping in view the type of security threats faced by cloud users today. For example, using AAA protocol and taking care of all the vulnerabilities to it.

2. **Access Management:** End users seek privacy and security of their data. Various intrusion detection systems, firewall, anti malware programs must be implemented at the cloud servers so as to prevent any malicious activity. For example, extensible Access Control Markup Language(XACML) can be used to control access to cloud applications.

3. **Data Protection:** Encryption tools, proper methods for authentication, tools for traffic monitoring and preventing various attacks such as DOS, wifi sniffing etc should be used to ensure security of data.

Apart from these techniques, various other specific measures need to be implemented at different levels to maintain privacy and integrity of data. These techniques should be designed for all three levels of the cloud infrastructure as well as the network through which data has to travel.

## VII. CONCLUSION AND FUTURE WORK

No matter how improved infrastructure or platform or the services cloud provides, until security related issues are addressed, complete confidence of users cannot be gained. As the number of mobile users increases, the data that is being shared through the cloud is definitely going to increase. In this paper, efforts have been put into studying the key areas where security issues need to be addressed. A lot more needs to be done as vulnerabilities go on increasing. Finding counter measures for each of these threats can be taken as future work. There always has to be a tradeoff between performance and security. In future, we can try to design such applications and protocols so that security is not compromised at the cost of enhanced performance.